\documentclass[11pt]{article}
\usepackage[T1]{fontenc}
\usepackage{times}
\usepackage{jeep}

\begin{document}

\title{When is a quantity additive, and when is it extensive?\thanks{
Contribution to the Proceedings of the International School and Conference
on Non-Extensive Thermodynamics and Physical Applications, Villasimius,
Sardegna, Italy, May 2001.}}
\author{Hugo Touchette\thanks{
E-mail addresses: htouc@cs.mcgill.ca, htouchet@alum.mit.edu (H. Touchette).} 
\\
\textit{Department of Physics and School of Computer Science,}\\
\textit{McGill University, Montr\'{e}al, Qu\'{e}bec, Canada H3A 2A7}}
\maketitle

\begin{abstract}
The difference between the terms additivity and extensivity, as well as
their respective negations, is critically analyzed and illustrated with a
few examples. The concepts of subadditivity, pseudo-additivity, and
pseudo-extensivity are also defined.
\end{abstract}

\section*{ }

To say that a given quantity (physical or mathematical) is additive and to
say that a quantity is extensive are two different affirmations which,
unfortunately, appear too often as undifferentiated in the physics
literature. The assimilation of the two terms is especially present in
studies related to the entropy measure of Tsallis \cite{tsallis1988}, and to
the so-called field of non-extensive thermostatistics or Tsallis' statistics
which is based on this measure of entropy \cite{abe2001}. In these studies,
it is not uncommon to see the words additivity and extensivity being used as
synonymous, and to read sentences such as ``...the appropriate framework to
describe non-extensive behavior is Tsallis' statistics, because of the
non-additivity property of Tsallis' entropy.'' But, how exactly is the
non-additivity property of Tsallis' entropy related to non-extensivity? Are
these concepts linked together simply because they are thought to mean the
same thing?

These questions are raised not with the intention of criticizing the results
related to non-extensive statistics; what is more important is the fact that
they point to a somewhat misleading and careless usage of scientific jargon.
That this carelessness persists would not by itself be so problematic, were
it not for the fact that the difference between additivity and extensivity
is at the very root of the issues raised by non-extensive statistics. For
this reason, it seems more than advisable to rehabilitate the proper meaning
of these two terms by reviewing their standard `textbook' definitions. This
is the purpose of the present contribution. Of course, in doing so, no new
results are likely to be conveyed; however, it can only be at our advantage
to repeat a number of concepts and definitions which are fundamental to
statistical physics. As Wittgenstein once wrote: ``what can be said at all,
can be said clearly.'' After reading this short review, it is hoped that
researchers will follow this advice, and will indeed be more perspicuous in
discussing any subject which connects itself to additivity and extensivity.

To begin this review, let us fix the context of the discussion. The
qualifiers `additive' and `extensive' apply to systems which are \textit{
composite} in the sense that they are composed of many components or
subsystems. A typical example of such composite systems is a gas composed of
many identical particles which may or may not interact with each others. If
we label these particles by the index $i=1,2,\ldots ,n$, and represent the
state of each of these particles by using the variable $x_i$, then the
possible interactions which may exist between the $n$ particles can be
modeled either in the form of a \textit{total energy} or \textit{Hamiltonian
function }$U(x^n)$ associated with the \textit{joint} state $
x^n=x_1x_2\ldots x_n$ of the complete system, or, more generally, by a 
\textit{joint probability distribution} $p(x^n)$ which gives the probability
that the particle 1 of the gas is in the state $x_1$, particle 2 in $x_2$,
and so on.

Now, suppose that we are interested in calculating or in measuring
physically a quantity (an observable) which is a function, say $Q(x^n)$, of
the joint state of a composite system. For this quantity, we may ask: can $
Q(x^n)$ be evaluated with the recourse of terms of order lower than $n$,
i.e., with terms like $Q(x^m)$, where $x^m$ is any $m$-particles \textit{sub}
-state of $x^n$ with $m<n$? In particular, can it be decomposed as the sum
of the $Q(x_i)$'s? The concept of additivity partly classifies, as follows,
the answers that these two questions can receive.

\begin{quote}
\textbf{Additivity.} A many-body (or joint) physical observable $Q(x^n)$ is
said to be \textit{additive} with respect to two subsystems with states $
x^m=x_1x_2\ldots x_m$ and $x^{n-m}=x_{m+1}x_{m+2}\ldots x_n$ if $
Q(x^n)=Q(x^m)+Q(x^{n-m})$.
\end{quote}

\noindent Obviously, this definition can be generalized to any partitions of
the joint state $x^n$ in two or more subsystems. As a result, we have the
following stronger version of additivity.

\begin{quote}
\textbf{Complete additivity.} $Q(x^n)$ is said to be \textit{completely
additive} if it is the sum of the one-particle (or marginal) contributions
to $Q$, i.e., if $Q(x^n)=\sum_{i=1}^nQ(x_i)$.
\end{quote}

A first example of a completely additive quantity which immediately comes to
mind is the total energy of a set of non-interacting bodies (e.g., the
particles of an ideal gas). `Non-interacting', in this case, can in fact be
taken to be equivalent to `additive energy', because, as we all know,
interactions appear in the expression of the total energy as extra
irreducible energy or potential terms which, in effect, allow a finite
number of bodies to exchange energy. These terms are \textit{irreducible} in
the sense that they are themselves non-additive.

In the case where $p(x^n)$ is given instead of $U(x^n)$, a similar
characterization of the properties of $p(x^n)$ is possible if one interprets
the logarithm of the inverse of this quantity as some kind of energy $
\mathcal{U}(x^n)$, or \textit{pseudo-energy} in order to distinguish it from
the `real' energy $U(x^n)$. This is the basis of the so-called thermodynamic
formalism of stochastic systems \cite{ruelle1978,beck1993,badii1997}. In
short, if $p(x^n)$ is \textit{factorizable} or \textit{separable}, i.e., if
it can be expressed as the product $p(x^n)=p(x_1)p(x_2)\cdots p(x_n)$ of
marginal probability distributions, then the corresponding pseudo-energy is
additive: 
\begin{equation}
\mathcal{U}(x^n)=-\ln p(x^n)=-\sum_{i=1}^n\ln p(x_i)=\sum_{i=1}^n\mathcal{U}
(x_i).  \label{eneadd1}
\end{equation}
On the other hand, when $p(x^n)$ is non-factorizable, $\mathcal{U}(x^n)$
becomes non-additive. Indeed, in this latter case, $p(x^n)$ can only be
written generically as a product of conditional probabilities of the form 
\begin{equation}
p(x^n)=p(x_1)p(x_2|x_1)p(x_3|x_1x_2)\cdots p(x_n|x^{n-1}),
\end{equation}
which makes obvious that Eq.(\ref{eneadd1}) is no more satisfied. In fact,
each conditional probability term $p(x_m|x^{m-1})$ above can be assimilated,
again by taking the logarithm of its inverse, to an interaction
pseudo-energy $\mathcal{U}(x^m)$ which enters in a non-additive fashion in
the expression of $\mathcal{U}(x^n)$. This makes the connection with the
properties of $U(x^n)$ more evident. (The connection is even more effective
when both $U(x^n)$ and $\mathcal{U}(x^n)$ can be defined and related to each
other, as in the case of thermodynamic systems. See Refs.\cite
{beck1993,badii1997} for more details.)

Another quantity which is also completely additive for independent systems,
and which is worth mentioning because of its great importance in statistical
mechanics, is the joint Boltzmann-Gibbs-Shannon (BGS) entropy of a sequence $
X^n=X_1X_2\ldots X_n$ of random variables: 
\begin{equation}
H(X^n)=-\sum_{x^n}p(x^n)\ln p(x^n).
\end{equation}
By inspecting the above expression, one can see that the joint entropy $
H(X^n)$ is nothing more than the expected total pseudo-energy (also called
bit-number \cite{beck1993}) over all joint states $x^n$. As a consequence,
one readily concludes that $H(X^n)$ is additive if and only if $p(x^n)$ is
factorizable. As an added property, it can also be shown that $H(X^n)$ is
strictly \textit{subadditive} otherwise, which means that $
H(X^n)<\sum_{i=1}^nH(X_i)$ when the random variables forming the state $X^n$
are correlated in one way or another \cite{cover1991}. At glance with the
BGS entropy, the joint version of the entropy measure of Tsallis, given by 
\begin{equation}
H_q(X^n)=\frac 1{q-1}\left( 1-\sum_{x^n}p(x^n)^q\right) ,
\end{equation}
is non-additive for independent random variables. For instance, it can be
verified that 
\begin{equation}
H_q(X_1,X_2)=H_q(X_1)+H_q(X_2)+(1-q)H_q(X_1)H_q(X_2)
\end{equation}
for $X_1$ and $X_2$ such that $p(x_1,x_2)=p(x_1)p(x_2)$ for all $x_1$ and $
x_2$ \cite{tsallis1988}. It is precisely because of this property that
Tsallis' entropy is referred to generically (but perhaps too generically) as
being non-additive.

Let us now turn our attention to another aspect of composite observables
which concerns their scaling properties with the number of subsystems. Given
again a function $Q(x^n)$ of the joint state $x^n$, we are interested in
determining if $Q(x^n)$ scales proportionally with $n$ or, at least,
asymptotically proportionally with $n$ when this number becomes large. If it
does, then $Q$ is termed \textit{extensive}. More precisely:

\begin{quote}
\textbf{Extensivity.} A joint observable $Q(x^n)$ is \textit{extensive} if
the $Q$\textit{-density}, defined by the ratio $Q(x^n)/n$, reaches a
constant in the limit $n\rightarrow \infty $. If random variables are used,
a criterion of convergence (e.g., convergence in probability, almost surely,
etc.) should also be provided in order to give a meaning to the limit.
\end{quote}

\noindent In relation to this definition, physicists commonly say that a
system possesses a \textit{thermodynamic limit} if its total energy and
total entropy are both extensive quantities \cite{beck1993,badii1997}.

At this point, it is important that the reader clearly distinguishes the
concept of extensivity from that of additivity. In general, extensivity does
not imply additivity, nor does additivity imply extensivity. (The latter
assertion is a common pitfall.) On the one hand, there exist joint
observables which are non-additive, but which are nonetheless extensive. The
joint BGS\ entropy of \textit{ergodic} sequences of correlated random
variables is often cited as an example of a quantity which is non-additive,
but which converges in density to a constant known as the \textit{entropy
rate}. (See \cite{badii1997,cover1991} for a definition of ergodic
stochastic processes, and \cite{cover1991} for a description of entropy
rates.) On the other hand, a quantity may well be additive without being
extensive. For instance, it is not so much difficult to construct a system
composed of independent particles characterized by a total additive energy $
U(x^n)$ which has no limits in $n$. But, admittedly, this would be a very
peculiar and exceptional example of an additive yet non-extensive quantity,
since, in most cases encountered in physics, additivity \textit{does} imply
extensivity. Furthermore, in many models of physical systems, the fact that
a quantity $Q$ is non-extensive is intimately related to the fact that it is
non-additive, for it is usually the non-additive part of this quantity which
divergences as $n\rightarrow \infty $, and which thus prevents the $Q$
-density to converge. Such a singularity occurs, for instance, when the
interaction energy or pseudo-energy between subsystems is \textit{long-range}
in the sense that it decays very slowly with the distance (if we are
concerned with particles in space) or with the neighborhood separation (as
for spin-lattice systems). The dimensionality of the system considered plays
also an important role in defining what `long-range' means (see, e.g., \cite
{ruelle1978,badii1997}).

By way of conclusion, note that it is sometimes possible to transform a
non-extensive quantity into an extensive one by rescaling it using a volume (
$n$) dependent factor or, simply, by dividing the quantity in question by
its nonlinear scaling law if it is characterized by such a law (Kac's
prescription \cite{ruffo2001}). The `renormalized' quantity which results
from this procedure is called \textit{pseudo-extensive} (see \cite
{tsallis1998} for a specific example). On a related note, a non-additive
quantity $Q(x^n)$ may be written in a \textit{pseudo-additive} way if one
can find another quantity $\tilde{Q}(x_i)$ such that 
\begin{equation}
Q(x^n)=\sum_{i=1}^n\tilde{Q}(x_i),
\end{equation}
with the additional requirement that $\tilde{Q}(x_i)=Q(x_i)$ for all $i$
whenever $Q(x^n)$ is additive. This last concept of pseudo-additivity is
probably not well-known to many; yet, it is effectively in use when one
expresses the total energy of a gas of electrons in a semiconductor as a sum
of free electron energies with a different renormalized or effective mass
factor. At the level of the entropy, a related procedure has also been
described recently by which the joint BGS entropy $H(X^n)$ of an ergodic
source of correlated random variables can be transformed in an
pseudo-additive fashion as the sum of the marginal Tsallis' entropies $
H_q(X_i)$, $i=1,2,\ldots ,n$, all assuming the same fixed value of $q\geq 1$ 
\cite{hugo2001}. This opens up, perhaps surprisingly, the possibility of
applying Tsallis' entropy to a wider class of systems than previously
thought, and, particularly, to the class of fully extensive systems which
have been hitherto neglected by the community of researchers working on
Tsallis' statistics.

\section*{Acknowledgments}

I wish to thank the organizers of the NEXT2001 Conference for their
application in making the conference a pleasant and fruitful meeting, and
for having provided the financial support which enabled me to cross the
Atlantic Ocean. This work was sponsored in part by NSERC (Canada), and FCAR
(Qu\'{e}bec).

\end{document}